\documentclass{aa}  

\usepackage{graphicx}
\usepackage{txfonts}
\usepackage{hyperref}
\hypersetup{colorlinks=true,linkcolor=blue,citecolor=blue,filecolor=blue,urlcolor=blue}
\usepackage{natbib}
\usepackage{soul}

\newcommand{\four}{4U~1812$-$12}

\newcommand{\ha}{H$\alpha$}

\newcommand{\lx}{$L_\mathrm{X}$}
\newcommand{\ledd}{$L_\mathrm{Edd}$}
\newcommand{\Nh}{$N_{\rm H}$}

\newcommand{\Msun}{\mathrm{M}_{\odot}}
\newcommand{\lum}{\mathrm{erg~s}^{-1}}
\newcommand{\flux}{\mathrm{erg~cm}^{-2}~\mathrm{s}^{-1}}

\newcommand{\nh}{$\mathrm{cm}^{-2}$}

\newcommand{\kms}{\mbox{${\rm km}\:{\rm s}^{-1}\:$}}

\begin{document}

   \title{Optical spectroscopy of \four: }
   \subtitle{an ultra-compact X-ray binary seen through an \ion{H}{ii} region}

   \author{M.~Armas~Padilla
          \inst{\ref{inst1},\ref{inst2}}
          \and
          T.~Mu\~noz-Darias \inst{\ref{inst1},\ref{inst2}}
          \and
          F.~Jim\'enez-Ibarra\inst{\ref{inst1},\ref{inst2}}
          \and
          J.~A.~Fern\'andez-Ontiveros\inst{\ref{inst3}}
          \and
          J.~Casares\inst{\ref{inst1},\ref{inst2}}
          \and
          M.~A.~P.~Torres\inst{\ref{inst1},\ref{inst2}}
          \and
          J.~Garc\'ia-Rojas\inst{\ref{inst1},\ref{inst2}}
          \and
          V.~A.~C\'uneo\inst{\ref{inst1},\ref{inst2}}
          \and
          N.~Degenaar\inst{\ref{inst4}}
          }

   \institute{Instituto de Astrof\'isica de Canarias (IAC), V\'ia L\'actea s/n, La Laguna 38205, S/C de Tenerife, Spain\label{inst1}\\
              \email{m.armaspadilla@iac.es}
         \and
             Departamento de Astrof\'isica, Universidad de La Laguna, La Laguna, E-38205, S/C de Tenerife, Spain\label{inst2}  
         \and
         Istituto di Astrofisica e Planetologia Spaziali (INAF--IAPS), Via Fosso del Cavaliere 100, I--00133 Roma, Italy \label{inst3} 
         \and
         Anton Pannekoek Institute for Astronomy, University of Amsterdam, Science Park 904, 1098 XH, Amsterdam, the Netherlands \label{inst4}    
             }


 
  \abstract 
  {
   The persistent, low-luminosity  neutron star X-ray binary \four\ is a potential member of the scarce family of ultra-compact systems. We performed deep photometric and spectroscopic optical observations with the 10.4~m Gran Telescopio Canarias in order to investigate the chemical composition of the accreted plasma, which is a proxy for the donor star class. We detect a faint optical counterpart (\textit{g}$\sim$25, \textit{r}$\sim$23) that is located in the background of the outskirts of the Sharpless~54 \ion{H}{ii} region, whose characteristic nebular lines superimpose on the X-ray binary spectrum.  Once this is corrected for, the actual source spectrum lacks hydrogen spectral features. In particular, the \ha\ emission line is not detected, with an upper limit (3$\sigma$) on the equivalent width of $<$1.3 \AA. Helium (\ion{He}{i}) lines are neither observed, albeit our constraints are not restrictive enough to properly test the presence of this element. We also provide stringent upper limits on the presence of emission lines from other elements, such as C and O, which are typically found in ultra-compact systems with C-O white dwarfs donors. The absence of hydrogen features, the persistent nature of the source at low luminosity, as well as the low optical to X-ray flux ratio confirm \four\ as a compelling ultra-compact X-ray binary candidate, for which we tentatively propose a He-rich donor based on  the optical spectrum and the detection of short thermonuclear X-ray bursts. In this framework, we discuss the possible orbital period of the system according to disc instability and evolutionary models.

  }


   \maketitle
%

\section{Introduction}

Ultra-compact X-ray binaries are systems of great interest for a number of reasons (e.g. stellar and binary evolution, gravitational wave detection, thermonuclear burning; \citealt{Nelemans2009,Tauris2018}). These X-ray binaries, which up to date have been found to harbour neutron star accretors, are tight in orbital periods shorter than $\sim$80~min where only a degenerate companion, such as a He star or white dwarf, can fit within the small Roche lobe \citep{Rappaport1982}. The current population of ultra-compact X-ray binaries is comprised by 14 systems (i.e. with reported orbital periods; see e.g. \citealt{Heinke2013}, and references therein; \citealt{Strohmayer2018}), despite $\sim$~10$^{5}$ of these sources are predicted to be hosted in our Galaxy \citep{VanHaaften2013}. We might be missing large numbers of these systems because of the observational challenges in measuring orbital periods of low-mass X-ray binaries in general \citep[LMXB;][]{Casares2017}, and the faintness of these systems, in particular. As a matter of fact, some indirect evidences used to identify ultra-compact candidates are based on their X-ray faintness, a direct consequence of their small accretion discs. According to the disc instability model, systems persistently accreting at very low X-ray luminosities should have short orbital periods, since smaller discs can be entirely ionised at lower accretion rates \citep{Lasota2001,IntZand2007}. Thus, ultra-compact systems could sustain a persistent behaviour when accreting at \lx$\lesssim10^{36}$~$\lum$, while systems with longer orbits have persistent X-ray luminosities of $\sim10^{37-38}$~$\lum$ \citep[see e.g.][]{Nelemans2010,Revnivtsev2012} . Likewise, small accretion discs result in lower optical to X-ray flux ratios because the disc region responsible for the X-ray (to optical) reprocessing is also smaller \citep{Paradijs1994}. Additional (also indirect) diagnostics are based on the degenerate nature of the donor star. These rely on the absence/presence of emission/absorption features in the source spectra, which provide hints on the chemical composition of the accreted material, and thus, on the nature of the companion star. We refer to \citet{Nelemans2004}; \citet{Baglio2016}; \citet{Santisteban2019} for examples of optical studies, \citet{Homer2002};  \citet{Tudor2018} of UV studies and e.g. \citet{Juett2003b}; \citet{Schulz2001}; \citet{ArmasPadilla2019b} for studies at X-ray energies.

The X-ray binary \four\ fulfils several of the above distinctive features. Since its discovery by the Uhuru mission \citep{Forman1976}, the source has consistently displayed a low (unabsorbed) X-ray flux of a few times 10$^{-10}~\flux$ \citep[e.g.,][]{Warwick1981, Wilson2003, Barret2003, Muno2005}. Numerous type I bursts have been reported, unveiling the neutron star nature of the compact object \citep[e.g.][]{Murakami1983, Tarana2006}. Many of these bursts showed photospheric radius expansion, which, assuming Eddington-limited luminosities, place the source at $\sim$~3.4 -- 4.6~kpc \citep{Cocchi2000, Jonker2004}. This implies a persistent X-ray luminosity of $\sim10^{36}$
~$\lum$ [$\sim$1 per cent of the Eddington luminosity (\ledd)]. In the same way, the \textit{R}=22.15 optical counterpart  corresponds to an absolute magnitude of $\sim$~3.6 -- 4.2 \citep{Bassa2006}. Thus, both the persistently X-ray low luminosity and the low optical to X-ray flux ratio make \four\ a promising ultra-compact X-ray binary candidate \citep{Bassa2006, IntZand2007}. 

In order to investigate further whether \four\ belongs to the ultra-compact class, we obtained deep optical spectroscopy and \ha\ imaging with the 10.4~m Gran Telescopio Canarias (GTC). Here, we present the results of this campaign, including the identification of an extended nebula in the sky region surrounding the system.

\begin{figure*}
\centering
\begin{center}
\includegraphics[keepaspectratio,width=15.0cm,  trim=2.0cm 0.0cm 0.0cm 0.0cm]{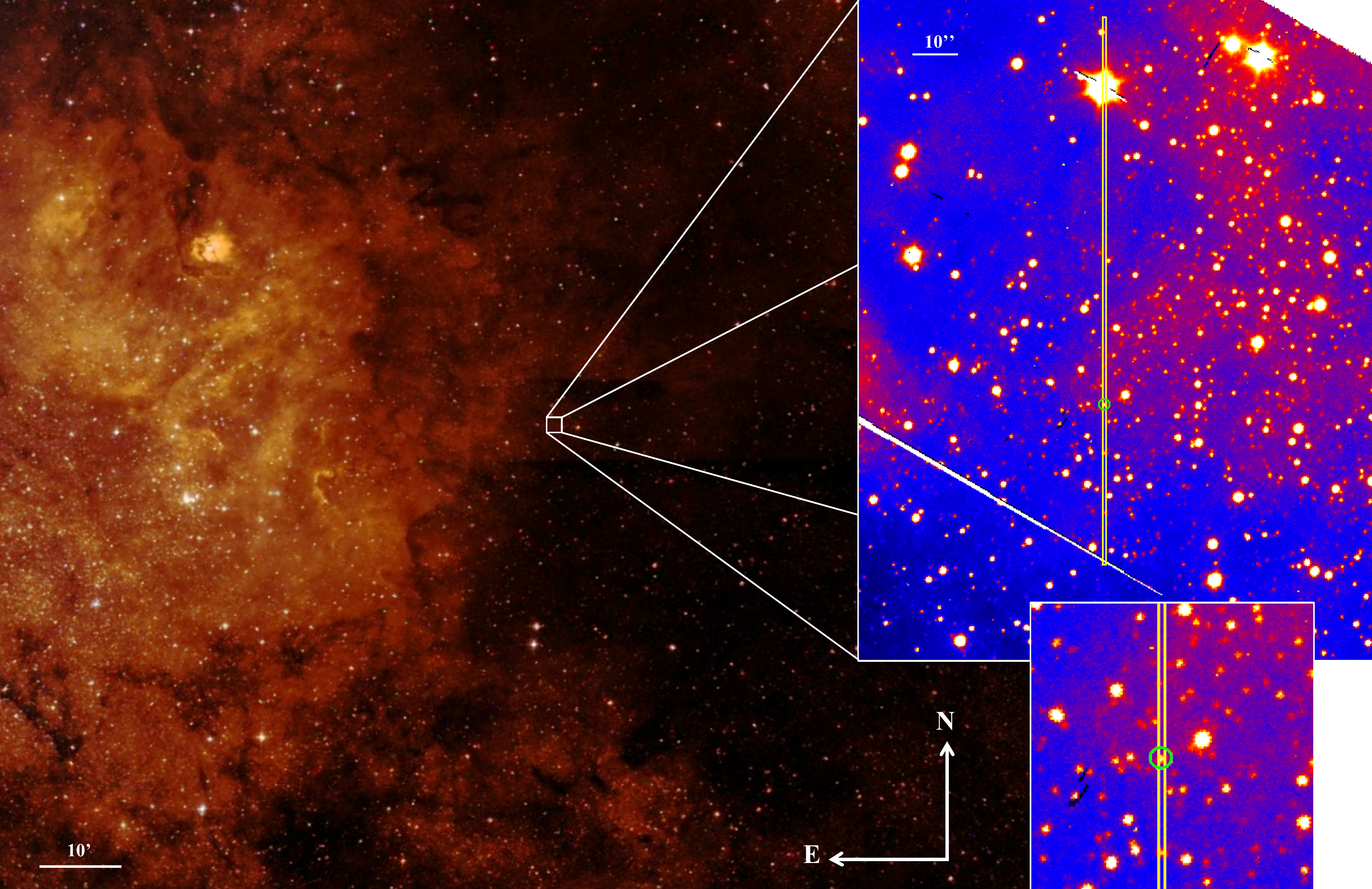}\\
\includegraphics[keepaspectratio,width=17.0 cm,  trim=2.0cm 0.0cm 0.0cm 0.0cm]{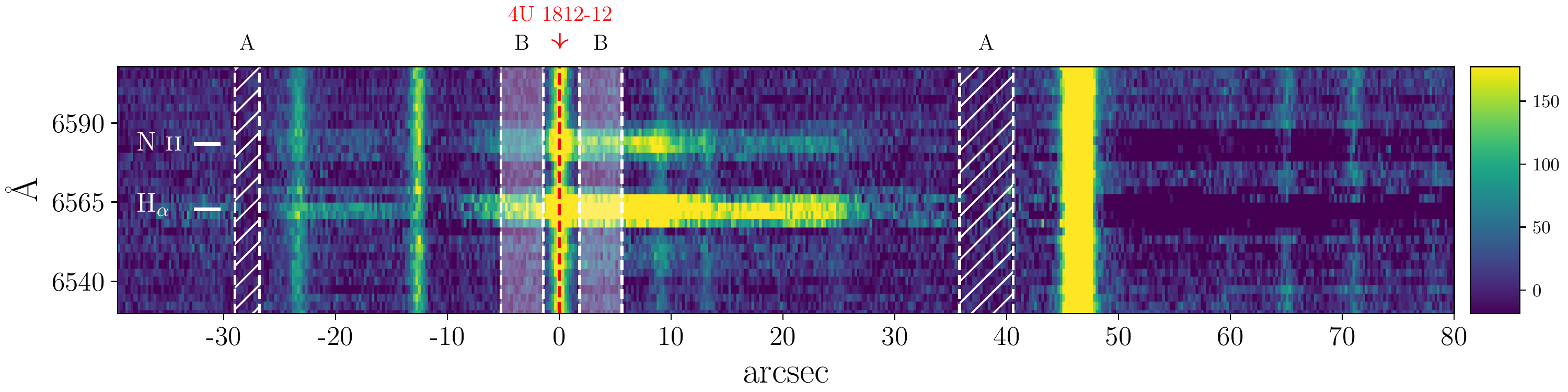}\\
\includegraphics[keepaspectratio,width=17.0cm,  trim=2.0cm 0.0cm 0.0cm 0.0cm]{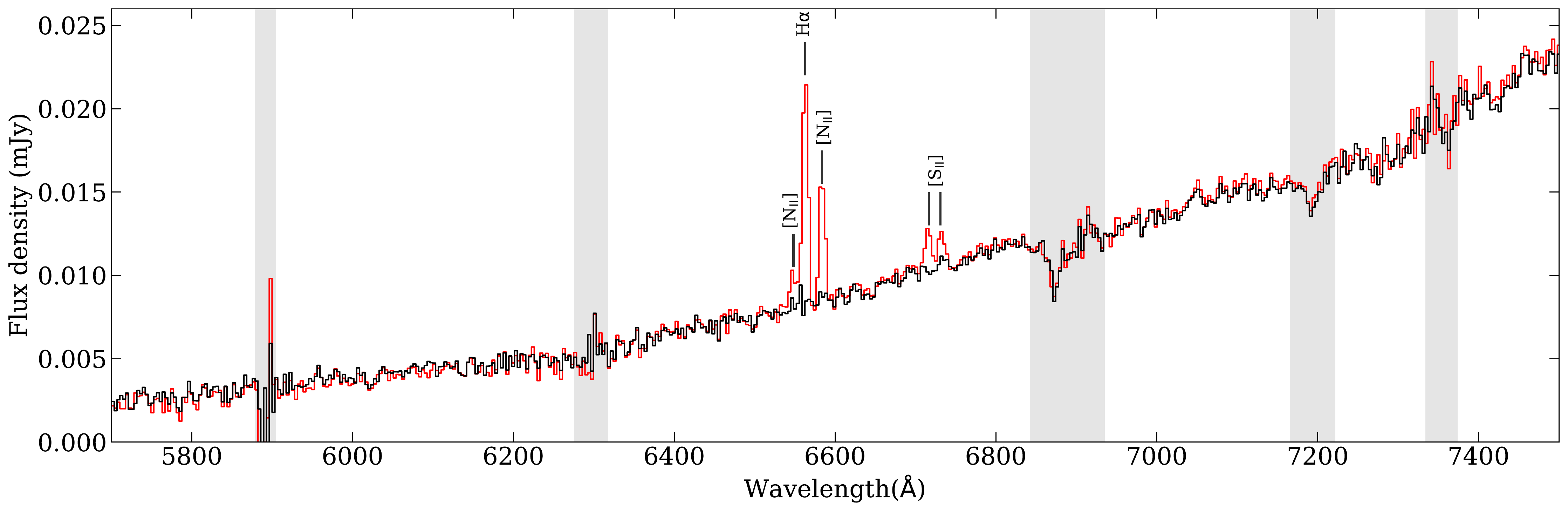}\\

\caption{\textit{Top panel}: DSS2 coloured image of 2$\degr \times$1.8$\degr$ including \four\ (white rectangle), which is located in the outskirts of the \ion{H}{ii} nebula Sharpless~54. A zoom-in of the target's region using our \ha\ narrow band imaging is shown together with the position of the slit in the spectroscopic data.  A 120~arcsec region covered by the slit along the spatial axis and corresponding to the 2-D spectrum (middle panel) is represented by a yellow rectangle, while the position of the system is marked by a green circle. For a better visualization, the closest region to the target is further zoomed. \textit{Middle panel}: Zoomed 2-D spectrum in the Halpha region, covering 120~arcsec in the spatial direction. The position of \four\ is marked by a dashed red line. The vertical, white stripes denote the different background regions considered for the overall (A) and nebula-free (B) spectra (see text). Intensity is indicated by a colour scale from 0 (deep blue) to 160 (bright yellow) counts.  \textit{Bottom panel}: Average spectra obtained by considering the `A' (red) and `B' (black) background regions. Grey-shaded bands indicate spectral regions affected by interstellar or telluric features. }
\label{fig:nebula}
\end{center}
\end{figure*}


\begin{figure*}$\phantom{!t}$
\centering
\begin{center}
\includegraphics[keepaspectratio, width=17.0cm, trim=2.0cm 0.0cm 0.0cm 0.0cm]{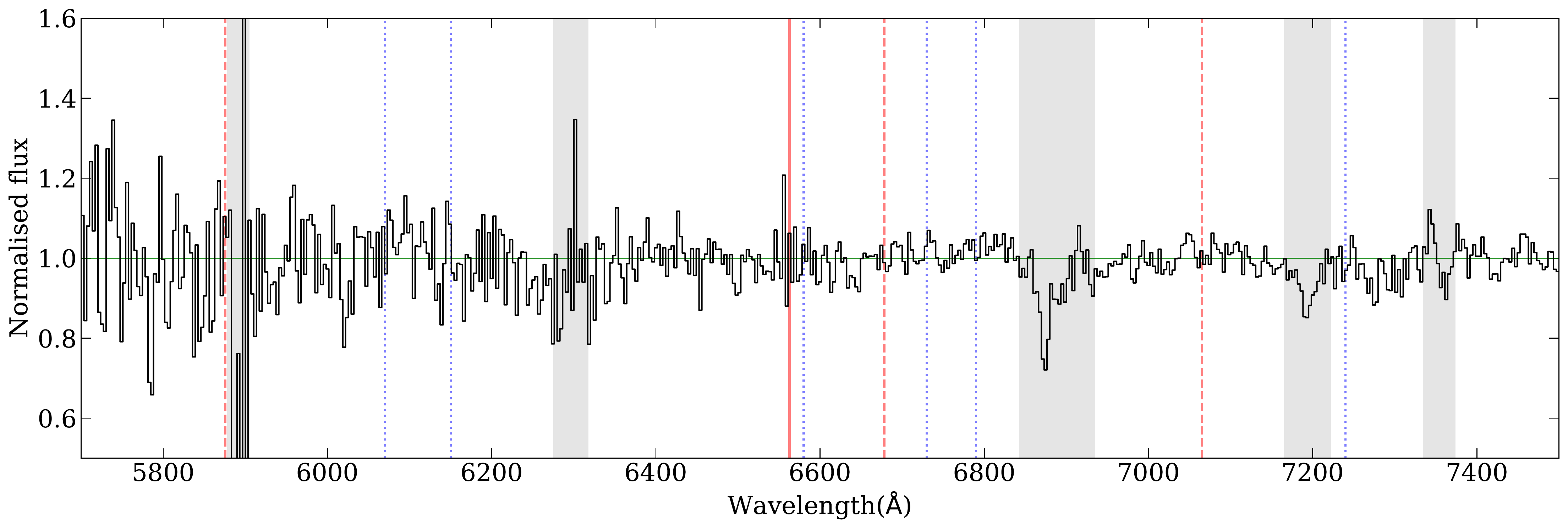}
\caption{Normalised spectrum of \four\ once the nebular contribution has been subtracted (i.e. using the B-labelled regions from Fig.\ref{fig:nebula} for background correction). The red vertical lines indicate the position of \ha\ (solid) and several He emission lines (dashed). The vertical dotted, blue lines indicate the approximate location of C and O blends detected in some ultra-compact systems \citep{Nelemans2004}. The grey-shaded bands flag the spectral regions affected by interstellar or telluric features.}
\label{fig:spec_core}
\end{center}
\end{figure*}


\section{Observations and data reduction}\label{sec:Obs}

We observed \four\ on 2018 June 13 with the Optical System for Imaging and low-Intermediate Resolution Integrated Spectroscopy \citep[OSIRIS,][]{Cepa2000} attached to the GTC at the Observatorio del Roque de Los Muchachos in La Palma, Spain. We used the grism R1000R  (2.62~\AA~pix$^{-1}$), which covers the spectral range 5100--10000~\AA. We took four spectra of 1400~s each using a slit-width of 1~arcsec, which provided a spectral resolution of 474$\pm$3~$\kms$ (measured as the full-width at half-maximum of a sky line at $\sim$6300~\AA). We reduced and combined the data using \textsc{iraf}\footnote{IRAF, the Image Reduction and Analysis Facility, is distributed by the National Optical Astronomy Observatory, which is operated by the Association of Universities for Research in Astronomy under cooperative agreement with the National Science Foundation.} tools and custom \textsc{python} routines, whereas the package \textsc{molly}\footnote{\url{http://deneb.astro.warwick.ac.uk/phsaap/software/molly/html/INDEX.html}} was used for the analysis.  

In addition, we obtained narrow band GTC/OSIRIS photometry ($3\times900$ sec) of the field (August 1, 2019) using the SHARDS filter U653/17, which is centred at 6530~$\AA$ (i.e. close to \ha) and has a bandpass of 160~$\AA$. We also took 120~sec images using the filters SDSS-\textit{g} and SDSS-\textit{r} for photometric measurements. Data were reduced using \textsc{astropy-ccdproc} routines \citep{Astropy2013,Astropy2018,ccdproc2017}.

\section{Analysis and results}\label{sec:Res}

Aperture photometry of the SDSS-\textit{g} and SDSS-\textit{r}  images was carried out using \textsc{photutils} \citep{photutils} and calibrated against field stars in the PanSTARRS catalogue \citep{PanStarrs}. The optical counterpart of \four\  is detected in both exposures, albeit it is significantly fainter in the blue filter (\textit{g}=25.3$\pm$0.3) than in the red band (\textit{r}=22.76$\pm$0.07). The latter is broadly consistent with \textit{R}=22.15 reported in \citet{Bassa2006}. This indicates that the source is significantly absorbed, in agreement wit \Nh$\sim$1.5$\times10^{22}$\nh derived in \citet{Tarana2006}. Using the standard \Nh -$A_{v}$ relation in \citet{Predehl1995} and the prescription of \citet{Yuan2013} to calculate reddening coefficients, this \Nh\ translates into $A_{g}\sim9$~mag and  $A_{r}\sim6.2$~mag. This is fairly consistent with the above reported magnitudes.
 
The trace of \four\ was clearly detected in the four spectra redward of 5700 \AA, which together with the strong telluric features present in the red part of the spectrum restrict our analysis to the spectral range 5700--7500 \AA. The 2-D spectra reveal the presence of extended emission in the \ha\ spectral region. A careful extraction using background regions not contaminated by this extended emission (labelled as `A'; middle panel in Fig. \ref{fig:nebula}), shows a strong and narrow \ha\ line accompanied by [\ion{N}{ii}]~${\lambda}$6548, 6584, as well as weaker [\ion{S}{ii}]~${\lambda}$6717, 6731 emission (see red spectrum in the bottom panel of Fig.~\ref{fig:nebula}). Unlike \ha, these forbidden transitions are not characteristic of LMXBs. A close look at the spatial profile of our spectra along the slit reveals that the lines  arise in an asymmetric extended region around the source (see middle panel in Fig.~\ref{fig:nebula}). The most intense emission region has a size of $\sim$~35~arcsec ($\sim$~25~arcsec from the source position to the North and $\sim$~10~arcsec to the South), although it expands further with a dimmer intensity. In a second step, we performed additional extractions considering background regions progressively closer to the trace of the target. As expected, this made the nebular emission weaker, and eventually to disappear. The bottom panel of Fig.~\ref{fig:nebula} shows the average spectrum obtained by considering the background regions labelled as `B' in the middle panel. These were chosen to minimise the presence of nebular [\ion{N}{ii}] features in the final spectrum. This also makes \ha\ to disappear (see sec. \ref{subsec:Ressource}). As can be noticed in Fig.~\ref{fig:nebula} (bottom panel) the two extractions, using sky regions `A' and `B',  yield remarkably similar continuum levels, with the exception of the spectral regions including the aforementioned nebular lines. This is consistent with the spectrum drawn in red being the sum of that of \four\ (i.e. the black one) and the contribution of a nebula in some characteristic transitions. For a more meaningful representation, both spectra were flux calibrated against the flux standard GD140 that was observed at the end of the night.      

The narrow-band imaging, top panel in Fig.~\ref{fig:nebula}, shows that \four\ is located in the outskirts of a region with a high density of diffuse gas. As matter of fact, the intensity profile of our 2-D spectrum (middle panel) matches the nebular structure covered by the slit in the spatial direction (top-right panel).
 
\subsection{The origin of the nebular emission}\label{subsec:ResNeb}
There is a variety of astrophysical objects, such as \ion{H}{ii} regions, planetary nebulae and supernova remnants, that naturally produce the nebular features that we have observed. In addition, some X-ray binaries have been associated with nebular structures produced by either photo-ionisation by high energy radiation \citep[e.g.][]{Cooke2007,Cooke2008} or shock-ionisation due to the impact of the binary jet on the interstellar medium \citep[e.g.][]{Russell2007, Wiersema2009}. 

In order to distinguish whether the surrounding diffuse emission is related to \four, we investigated some characteristic emission line ratios commonly used as diagnostics of shock-ionised or photo-ionised gas. In particular, we use the \citet{Sabbadin1976} diagnostic diagram, that uses the spectral lines that are more prominent in our spectra (\ha/[\ion{N}{ii}] vs. \ha/[\ion{S}{ii}]). We obtain \ha/[\ion{N}{ii}]$_{(6548+6584)}$=1.7$\pm$0.1 and  \ha/[\ion{S}{ii}]$_{(6717+6731)}$=4.1$\pm$0.4, values that comfortably sit on the zone occupied by \ion{H}{ii} regions (see fig. 10 in \citealt{Ottl2014}; see also \citealt{Magrini2003}). Exploring the location of our source, we tentatively identify the diffuse gas surrounding \four\ as being part of the large \ion{H}{ii} emission nebula Sharpless~54 (Sh2-54) placed eastward from the target. This region is ionised by the open cluster NGC~6604, located in the Serpens constellation, at $\sim$1.5~kpc (\citealt{Zucker2020}; see Fig.~\ref{fig:nebula}).  Thus, the nebular emission is most likely in the foreground of \four, which is located at $\sim$ 4 kpc from the Earth.  

\subsection{The optical spectrum of \four}\label{subsec:Ressource}
We inspected the spectrum of \four\ searching for features indicative of the chemical composition of the accreted material, and thus of the nature of the companion star. Neither H nor He emission lines are detected (see Fig.~\ref{fig:spec_core}). In Table \ref{table:EW} we list 3$\sigma$ upper limits on the equivalent width (EW) of \ha\ and some \ion{He}{i} lines, which are characteristic of LMXBs with H-rich companion stars \citep[e.g.][]{Charles2006,MataSanchez2018}.  These were calculated using 1500~\kms wide spectral regions, which is a standard value for the full width at half maximum of \ha\ in LMXBs \citep{Casares2015}, centred at the rest wavelength of each transition. 

Similarly, we also investigated the presence of C and O features, which have been detected in the optical spectra of some (candidate) ultra-compact X-ray binaries favouring C-O white dwarf companions \citep[see e.g.][]{Nelemans2004, Werner2006}.  None of these features are significantly detected (3$\sigma$) in our spectra [see Table~\ref{table:EW} for upper limits on the EW for the most prominent blends of C and O lines \citep{Nelemans2004,IntZand2008}]. 

\begin{table}
\caption{Upper limits ($3\sigma$) on the EW of some of the most prominent lines of classic LMXBs (i.e. with H-rich donors) and C-O blends detected in ultra-compact X-ray binaries \citep[see][]{Nelemans2004}.}
\label{table:EW}      
\centering
\begin{tabular}{c c c } 
\hline\hline  
Line & $\lambda_{\rm c}$ ($\AA$)   & EW ($\AA$) 	\\				 
\hline 
\ion{He}{i}     & 5875.6 	& $<$2.8  \\ 	
\ha\            & 6562.8    & $<$1.3 \\      
\ion{He}{i}     & 6678.1   	& $<$1.1 \\	    
\ion{He}{i}     & 7065.2    & $<$0.9   \\    
\hline\hline                    
							
Blend 		& Spectral range\tablefootmark{a} ($\AA$)    & EW ($\AA$)\\ 
\hline
\ion{C}{ii} & 6040--6120 &$<$4.6\\
\ion{C}{ii}--\ion{C}{iii}  & 6130--6180 & $<$3.3\\
\ion{C}{ii}--\ion{O}{ii}  & 6550--6600 & $<$1.5\\
\ion{C}{iii}--\ion{O}{ii} & 6700--6760 & $<$1.3\\
\ion{C}{ii}--\ion{C}{iii} & 6760--6820 & $<$1.3\\
\ion{C}{ii} & 7210--7260 & $<$1.1\\

\hline
\end{tabular}
\tablefoot{   
   \tablefoottext{a}{EWs calculated in the spectral ranges used in \citet{Nelemans2004}.}
   }
\end{table}

\section{Discussion}
The neutron star X-ray binary \four\ has been recursively detected at $\sim10^{36}~\lum$ in every X-ray observation performed since its discovery 50 years ago. This persistent activity at low luminosity guided \citet{IntZand2007} to propose an ultra-compact orbit for the source. Likewise, \citet{Bassa2006} suggested a short orbital period on the basis of the low optical to X-ray flux ratio \citep{Paradijs1994}. Here, we reported on deep, optical spectroscopy aiming at identifying the companion star class and further test the ultra-compact nature of the system. 

Our narrow-band imaging shows that \four\ is located in a patch of the sky with a high density of diffuse gas. The optical spectrum revealed conspicuous \ha, [\ion{N}{ii}] and [\ion{S}{ii}] emission lines, similar to those typically found in \ion{H}{ii} regions. The flux ratios derived from these features strongly support this association. Thus, we tentatively identify this diffuse gas region surronding the target as being part of the nearby \ion{H}{ii} nebula Sharpless~54 located at $\sim$1.5~kpc \citep{Zucker2020} in the Serpens constellation. Considering the distance of $\sim$4~kpc inferred from the detection of Eddington-limited thermonuclear bursts, we conclude that the nebular emission is in the foreground of \four\ and therefore that both astrophysical objects are unrelated. 

Once the nebular contribution is subtracted (see Section \ref{sec:Res}), the spectrum of \four\ lacks any of the strong H and He features typically found in LMXBs,  supporting the ultra-compact nature proposed for the system. On the one hand, in classic LMXBs (i.e., with H-rich, low mass companion stars) \ha\ is typically the most prominent optical emission line. Its EW is anti-correlated with the continuum luminosity (X-ray and optical), with typical values of EW$>$10~$\AA$ for X-ray luminosities  \lx$\lesssim10^{36}~\lum$ \citep{Fender2009a}.  This is significantly larger than our $\sim$1.3~\AA\ upper limit.  On the other hand, \ion{He}{i} features are much weaker in LMXBs, with \ion{He}{i} at 5876 \AA\ being typically the strongest line. Unfortunately, this transition is in the blue part of our spectral range, where the signal-to-noise ratio (SNR) is poorer. Thus, our upper limit on the EW ($<$2.8~\AA) is not particularly constraining. Likewise, our more stringent upper limits on the redder \ion{He}{i} transitions do not rule out the presence of He in the accretion disc. As a matter of fact, theoretical modelling of hot discs in ultra-compact systems favour the formation of  \ion{He}{ii} lines above those of  \ion{He}{i} \citep{Werner2006}. This is consistent with the modelling by \citet{Nelemans2006}, which predict the presence of strong  \ion{He}{ii} $\lambda$4686 and weaker  \ion{He}{i} $\lambda$5876, with the remaining He transitions being too weak to be detected unless very high SNR data are achieved.  Our deep photometry (\textit{g}$\sim$25) indicates that testing the presence of \ion{He}{ii} $\lambda$4686 in the spectrum of \four\ will not be feasible in the near future.

A number of confirmed ultra-compact systems also show featureless optical spectra. Deep Gemini spectroscopy of the $\sim38$~min orbital period LMXB IGR~J17062$-$6143 \citep{Strohmayer2018} reveal a blue continuum with no emission lines \citep{Santisteban2019}. Also, the ultra-compact system 2S~0918$-$549 ($P_{\rm orb}$=17.4~min,  \citealt{Zhong2011}) and the candidate A~1246-58 do not show significant features in Very Large Telescope spectra \citep{Nelemans2004, Nelemans2010,IntZand2008}. However, some ultra-compact systems did show emission lines of elements other than H. In particular, the detection of C and O emission features in the optical spectra of 4U~0614+091,  4U~1626$-$67 and  4U~1543$-$624 evidences the presence of metal-rich material in the accretion disc, suggesting a C–O white dwarf donor for these sources \citep{Nelemans2004,Nelemans2006,Werner2006,Baglio2014}. These emission lines are not present in our GTC spectrum. In particular, for the C-O blends between 6500 and 7300~\AA\ we derive 3$\sigma$ upper limits ($<$1.1--1.5~\AA) which are significantly smaller than the 1.5--4~\AA\ EWs reported in the aforementioned works. Likewise, the presence of \ion{He}{ii} and \ion{N}{iii} emission lines in the spectrum of the dipper system 4U~1916$-$05 provided the first direct evidence of a He-rich companion star in an ultra-compact X-ray binary \citep{Nelemans2006}.  This scenario is supported by the detection of frequent short bursts episodes, which, in this case, would be ignited by pure (or almost pure) He accreted fuel \citep[see e.g.,][]{Galloway2008}. Finally, we also note that the transient ultra-compact candidate 1RXS~J180408.9$-$342058 \citep{Baglio2016, Degenaar2016} shows a hint of \ion{He}{ii} emission at 4686~\AA. Based on this, \citet{Baglio2016} proposed a He white dwarf companion in a $\sim$40~min orbit.

As a matter of fact, the properties of thermonuclear X-ray bursts, such as duration, recurrence time and radiated energy, provide hints on the composition of the accreted fuel.  As a rule of thumb, regular (short, $\lesssim$ 1~min) X-ray bursts are fuelled by He or a mixture of H and He, while intermediate and long bursts (several minutes to hours) are powered by He and C burning, respectively \citep[see e.g.][]{Cumming2003, Falanga2008, Galloway2010}.  In this regard, \four\ has displayed numerous regular bursts \citep{Murakami1983,Cocchi2000,Tarana2006, Galloway2020}. Assuming the premise of the ultra-compact nature, these would be fuelled by pure He, pointing therefore to a He-rich donor, similarly to 4U~1916$-$05. Interestingly, the properties of \four\ resemble to some extend to those displayed by 4U~1916$-$05: a persistent X-ray luminosity of $\sim$0.01~\ledd, a low optical to X-ray luminosity ratio, the aforementioned episodes of short thermonuclear bursts and the lack of H lines in the optical spectra. As discussed above, the faintness of the source in the blue (\textit{g}$\sim$25) prevented us from testing the presence of the distinctive \ion{He}{ii}--\ion{N}{iii} lines observed in 4U~1916$-$05, which are located at 4000-5000~$\AA$ \citep{Nelemans2006}. Therefore, we can not rule out that a similar scenario might apply to \four.

This would not be in conflict with the disc instability model for He accretion discs. Following \citet{Coriat2012} we derive\footnote{We assumed an average  2-10~keV continuum X-ray unabsorbed flux of 3.8$\times$10$^{-10}$~$\flux$ \citep{Cocchi2000}, a bolometric correction of 2.9 \citep{IntZand2007} and a distance of 4.6~kpc for which an accretion of helium-rich material was assumed \citep{Jonker2004}.} a persistent mass-transfer rate of $\sim$5$\times$10$^{-10}$~$\Msun~\mathrm{yr}^{-1}$, which sits above the (irradiated He disc) instability threshold for orbital periods shorter than $\sim$~25~min \citep{Lasota2008}.  In addition, considering evolutionary tracks for donor stars in ultra-compact binaries, the above mass accretion rate translates into an orbital period of $\sim$~20~min for a He white dwarf donor \citep{Deloye2003, Sengar2017}. Using the same approach the orbital period would become $\gtrsim$~35~min  for a He star companion. However, the latter longer orbital period would imply a transient behaviour according to the instability model for He discs \citep{Heinke2013,Hameury2016}.

It should be noted, however, that in some cases is difficult to reconcile all the observables into a unified picture. For instance, the intermediate thermonuclear bursts displayed by 4U~0614+091 suggest a He-rich companion star, while its optical spectrum points to a C-O rich accretion disc \citep{Kuulkers2010}. Likewise, a He-rich companion was proposed for the ultra-compact 2S~0918$-$549 based on the detection of intermediate bursts \citep{IntZand2005b}, while the disc instability models favours a C-O white dwarf companion in order to account for its persistently low luminosity \citep{Heinke2013}. Spallation reactions \citep{Bildsten1992} have been invoked as a possible solution for this problem \citep{Juett2003b, Nelemans2004}. However, this scenario presents several caveats as discussed in \citet[][]{IntZand2005b} and \citet{Heinke2013}.

\section{Conclusions}\label{sec:conclu}
We have presented optical (GTC-10.4m) spectroscopy and \ha\ imaging of the X-ray binary \four, which is found to be located in the background of the \ion{H}{ii} region Sharpless~54. The source spectrum is featureless, not showing any of the emission lines typically displayed by LMXBs with H-rich companion stars. In particular, the absence of \ha\ emission strongly suggests an evolved donor star. We conclude that the lack of H features reported here, together with the persistently dim luminosity and low optical to X-ray flux ratio, strongly endorse the ultra-compact nature of \four. We suggest the companions star is He rich,  based on the optical spectrum and the short thermonuclear type-I bursts displayed by the system.

\begin{acknowledgements}
We acknowledge support from the State Research Agency (AEI) of the Spanish Ministry of Science, Innovation and Universities (MCIU) and the European Regional Development Fund (FEDER) under grant AYA2017-83216-P and AYA2017-83383-P. TMD and MAPT acknowledges support via a Ram\'on y Cajal Fellowship (RYC-2015-18148 and RYC-2015-17854, respectively). JAFO acknowledges financial support by the Agenzia Spaziale Italiana (ASI) under the research contract 2018-31-HH.0. JG-R acknowledges support from an Advanced Fellowship from the Severo Ochoa excellence program (SEV-2015-0548). JG-R also acknowledges support under grant P/308614 financed by funds transferred from the Spanish Ministry of Science, Innovation and Universities, charged to the General State Budgets and with funds transferred from the General Budgets of the Autonomous Community of the Canary Islands by the MCIU. ND is supported by a Vidi grant from the Netherlands Organisation for Scientific Research (NWO). Based on observations made with the Gran Telescopio Canarias (GTC), installed in the Spanish Observatorio del Roque de los Muchachos of the Instituto de Astrof\'isica de Canarias, in the island of La Palma. This work is partly based on data obtained with the SHARDS filter set, purchased by Universidad Complutense de Madrid (UCM). SHARDS was funded by the Spanish Government through grant AYA2012-31277. MOLLY software developed by T. R. Marsh is gratefully acknowledged.

\end{acknowledgements}

%
%

\bibliographystyle{aa} 
\bibliography{4U1812-12_arx.bbl}

\end{document}